\documentclass[square]{ws-procs11x85}
\usepackage{balance} 

\def\Journal#1#2#3#4{{#1} {\bf #2}, #3 (#4)}

\newcommand{\met}{\hbox{E\kern-0.5em\lower-0.1ex\hbox{/}}_T}
\newcommand{\ltsima} {$\; \buildrel < \over \sim \;$}
\newcommand{\simlt}  {\lower.5ex\hbox{\ltsima}}            
\newcommand{\gtsima} {$\; \buildrel > \over \sim \;$}
\newcommand{\simgt}  {\lower.5ex\hbox{\gtsima}}            

\begin{document}

\newcommand{\lsi}   {LS~I~+61~303}

\twocolumn[
\title{VHE Gamma-rays from Galactic X-ray Binary Systems}

\author{J.M. Paredes}

\address{Departament d'Astronomia i Meteorologia and Institut de Ci\`encies del Cosmos (ICC), Universitat de Barcelona (UB/IEEC), Mart\'{\i} i 
Franqu\`es 1,
08028 Barcelona, Spain\\E-mail: jmparedes@ub.edu}


\begin{abstract}
The detection of TeV  gamma-rays from LS~5039 and the binary pulsar PSR B1259$-$63 by HESS, and from \lsi\ and the stellar-mass black hole Cygnus~X-1 by MAGIC, provides a clear evidence of 
very efficient acceleration of particles to multi-TeV energies in X-ray binaries. These 
observations demonstrate the richness of non-thermal phenomena in compact galactic objects 
containing relativistic outflows or winds produced near black holes and neutron stars. I 
review here some of the main observational results on very high energy (VHE) $\gamma$-ray emission from 
X-ray binaries, as 
well as some of the proposed scenarios to explain the production of VHE $\gamma$-rays. I put special emphasis on the flare TeV emission, suggesting that the flaring activity might be a common phenomena in X-ray binaries.
\end{abstract}
\vskip12pt  
]

\bodymatter

\section{Introduction}
An X-ray binary (XB) is a binary system containing a compact object, neutron star (NS) or 
a stellar-mass black hole (BH), emitting in X-rays. In most cases, the X-ray emission is a consequence of the accretion of matter from the companion star.   Up to now, 299 accreting X-ray 
binaries have been catalogued in the Galaxy, being 114 High Mass X-ray Binaries (HMXBs) \cite{liu06}
and 185 Low Mass X-ray Binaries (LMXBs) \cite{liu07}. In the HMXBs, the optical companion has spectral type O or B and the mass transfer occurs via decretion disc (Be 
stars) or via strong wind or Roche-lobe overflow. In the LMXBs, the optical companion has 
spectral type later than B and the mass transfer is via Roche-lobe overflow. Among the 299 
known XBs, 65 of them (9 HMXB and 56 LMXB) are radio emitting sources. Some ($\sim$20)  of these radio 
emitting sources show a relativistic jet and are called microquasars (see 
Refs.~\refcite{mirabel99} and \refcite{fender06} for reviews on microquasars).

Some of these sources were observed with the first generation of TeV instruments. The source most widely observed was the HMXB Cygnus X-3 and although there were claims of its detection, it has not been considered as a 
positive one by the astronomical community because it has never been confirmed by the new generation of Cherenkov telescopes and the instrumentation at this epoch was limited.
Another monitored source was the LMXB  microquasar GRS 1915+105, which was the first superluminal source detected in the Galaxy \cite{mirabel94}. The HEGRA experiment
detected a flux of the order of 0.25 Crab from GRS 1915+105
during the period May-July 1996 when the source was in an active state \cite{aharonian98}. This source has also been observed
with Whipple, obtaining a 3.1$\sigma$ significance \cite{rovero02}. 
Although these sources have been observed with the new generation of Cherenkov telescopes and they remain undetected up to now, other binaries have been detected at TeV energies.   

Nowadays, the VHE $\gamma$-ray sky map contains four XBs, namely PSR~B1259$-$63  \cite{aharonian05a}, 
\lsi\ \cite{albert06}, LS~5039 \cite{aharonian05b} and Cygnus~X-1 \cite{albert07}. Whereas the optical companion in all these sources is well known, a massive O or B star, the nature of the compact companion is unknown in two of them. While in the case of PSR~B1259$-$63 it is clear that the compact
object is a rapidly spinning non-accreting NS and the TeV emission is
powered by rotational energy \cite{johnston92,aharonian05a}, and in the case of Cygnus~X-1 it is an accreting stellar-mass BH, in LS~5039 and \lsi\ the uncertainties in the determination of the inclination of the system \cite{casares05a,casares05b} precludes to fix the mass of the compact object and, therefore, to know if it is a BH or a NS.

In this work, I report in Section \ref{binaries} the main characteristics describing each of the four X-ray binaries detected at TeV energies up to now. In Section \ref{scenarios} I present some scenarios for high energy (HE) and VHE $\gamma$-ray production. Finally, in Section \ref{discussion} a brief discussion and summary is given.

\section{Binary TeV sources - BTV}\label{binaries}
Here follow some of the main characteristics of these binary TeV sources (BTV). Some of these characteristics are shared while other are unique and make each of these sources particular. Some properties are summarized in Table~\ref{aba:tbl1}.

\begin{table*}[t]
\tbl{The four X-ray binaries that are TeV emitters.}
{
\begin{tabular}{@{}ccccc@{}}
\toprule
{Parameters} & {PSR~B1259$-$63} & {LS I +61 303} & {LS 5039} & {Cygnus~X-1} \\
\colrule
System Type  & B2Ve+NS & B0Ve+NS?  & O6.5V+BH? & O9.7Iab+ BH \\[3pt]
Distance (kpc) & 1.5 & 2.0$\pm$0.2             & 2.5 & 2.2$\pm$0.2 \\[3pt]
Orbital Period (d)  & 1237 & 26.5  & 3.9 & 5.6 \\[3pt]
$M_{\rm compact}$ (M$_{\odot}$)  & 1.4   & 1--4  & 1.4--5  & 20$\pm$5 \\[3pt]   Eccentricity & 0.87 & 0.72      & 0.35 & $\sim$ 0 \\[3pt]
Inclination & 36 &  $30\pm 20$     & 20? & $33\pm 5$\\[3pt]
Periastron-apastron (AU) & 0.7--10 &  0.1--0.7  & 0.1--0.2 & 0.2\\[3pt]
Activity Radio & Periodic (48 ms and 3.4 yr)  &  Periodic (26.5 d and 4 yr)
   & persistent & persistent \\[3pt]
Radio Structure (AU) & $\simlt$2000 & Jet-like (10--700)  & Jet (10--$10^{3}$)  & Jet
(40) + Ring\\[3pt]
$L_{\rm radio(0.1-100GHz)}$ (erg s$^{-1}$) & $(0.02-0.3)\times10^{31}$ $^{\rm (*)}$& $(1-17)\times10^{31} $ &$1\times10^{31}$   & $0.3\times10^{31}$ \\[3pt]
$L_{\rm X(1-10keV)}$ (erg s$^{-1}$) & $(0.3-6)\times10^{33}$  & $(3-9)\times10^{33}$  & $(5-50)\times10^{33}$  & $1\times10^{37}$  \\[3pt]
$L_{\rm VHE}$ (erg s$^{-1}$) & $2.3\times10^{33}$$^{\rm (a)}$  & $8\times10^{33}$$^{\rm (a)}$  & $7.8\times10^{33}$$^{\rm (b)}$  & $12\times10^{33}$$^{\rm (a)}$  \\[3pt]
$\Gamma_{\rm VHE}$ & $2.7\pm0.2$ & $2.6\pm0.2$  & $2.06\pm0.05$ & $3.2\pm0.6$  \\[3pt]
Remarks & -- & 3EG~J0241+6103   & 3EG~J1824$-$1514 & --\\[3pt]
\botrule
\end{tabular}
\label{aba:tbl1}
}
{\small
$^{\rm (*)}$ Unpulsed radio emission; $^{\rm (a)}$ 0.2 $<$E $<$ 10 TeV; $^{\rm (b)}$ Time averaged luminosity.
}
\end{table*}

\subsection{PSR~B1259$-$63}
PSR B1259-63 / SS 2883 is a binary system containing a B2Ve donor and a 47.7 ms radio pulsar orbiting it every 3.4 years, in a very eccentric orbit with e=0.87. 
No radio pulses are observed when the NS is behind the circumstellar decretion disk 
(free-free absorption). VHE gamma-rays are detected when the NS is close to 
periastron or crosses the disk \cite{aharonian05a}. The TeV light-curve shows significant variability and a behavior not predicted by previously available models. The observed time-averaged energy spectrum can be fitted with a power law with a photon index $\Gamma_{\rm VHE}$ = $2.7\pm0.2_{\rm stat}\pm0.2_{\rm sys}$ 

Different models have been proposed recently to try to explain these observations. It is shown by Ref.~\refcite{khangulyan07} that the
observed TeV light-curve can be explained by the
inverse Compton model under certain assumptions, although the large  uncertainties, as well as the relatively narrow energy band of the available TeV data do not allow to put robust
constraints on model parameters. In Ref.~\refcite{neronov07},
a hadronic scenario is proposed where the TeV 
light-curve, and radio/X-ray light-curves, can be produced by the collisions of high energy protons accelerated by the pulsar with the circumstellar disk. In this scenario the VHE $\gamma$-rays are produced in the decays of secondary $\pi^{0}$, while radio and X-ray emission are synchrotron and inverse Compton emission produced by low-energy ($<$ 100 MeV) electrons from the decays of secondary $\pi^{\pm}$. 
Coordinated observations in the keV, MeV, GeV and TeV bands are very important to constrain the models, especially at the epochs close to the periastron and disk passages.

\subsection{\lsi}
\lsi\ is a HMXB system that shows periodic non-thermal radio
outbursts on average every $P_{\rm orb}$=26.4960~d \cite{gregory02}. The
system is composed of a rapidly rotating early type B0\,Ve star with a stable
equatorial decretion disk and mass loss, and a compact object with a mass
between 1 and 4 $M_{\odot}$ orbiting it every $\sim$26.5~d in an eccentric orbit \cite{casares05a}. The periastron takes place at phase 0.23$\pm$0.02. Spectral line radio observations give a distance of 2.0$\pm$0.2~kpc
\cite{frail91}. The maximum of the radio outbursts varies between phase 0.45 and 0.95. X-ray
outbursts, starting around phase 0.4 and lasting up to phase 0.6, have also been detected (see Ref.~\refcite{harrison00} and references therein). Orbital X-ray periodicity has also been found using RXTE/ASM data \cite{paredes97}, which currently reveal a broad maximum covering phases 0.4–-0.6. Similar results have recently
been obtained at higher energies with INTEGRAL data \cite{hermsen06}. 
The X-ray spectrum can be fitted with an absorbed
power-law with values in the ranges $N_{\rm
H}$=0.45--0.65$\times10^{22}$~cm$^{-2}$ and $\Gamma$=1.5--1.9 \cite{sidoli06,chernyakova06}. Recent {\it
Chandra} observations have reported  possible evidence of an extended X-ray emission 
from the source \cite{paredes07}.

\begin{figure}[t]
\center
\centerline{\psfig{figure=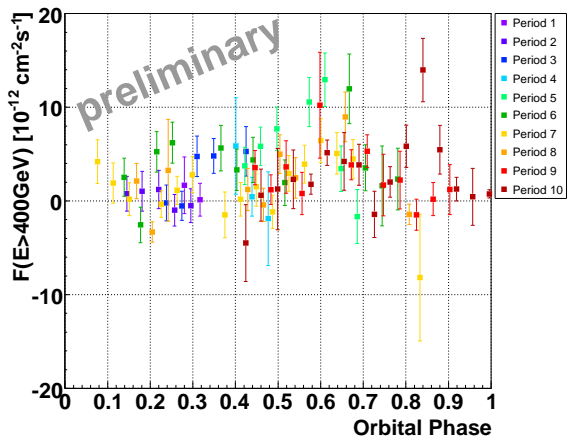, angle=-90, width=7.5truecm}}
\caption{MAGIC measured fluxes from \lsi\ as a function of orbital phase \cite{sidro07}.\label{fig:LSI}}
\end{figure}

\lsi\ is also spatially coincident with 3EG~J0241+6103, a HE ($>$ 100 MeV) source detected by EGRET \cite{kniffen97}. 
The MAGIC measurements showed that the VHE $\gamma$-ray emission from \lsi\
is variable and possibly periodic, with maximum flux $\sim$16\% that of the Crab Nebula 
detected around phase 0.6 \cite{albert06}. The VHE spectrum is fitted reasonably well by a power law: $dN/(dA dt dE) = (2.7\pm0.4\pm0.8)\times10^{-12}E$$^{(-2.6\pm0.2\pm0.2)}$ cm$^{-2}$ s$^{-1}$ TeV$^{-1}$. A preliminary analysis of new data, which confirms 
the former results, has been presented by Ref.~\refcite{sidro07}, and is shown in Fig.~\ref{fig:LSI}. In addition, the orbital TeV variability has been confirmed by the VERITAS Collaboration \cite{maier07}.

The discovery of apparently relativistic radio emitting jets in \lsi\ allowed to classify this XB as a microquasar \cite{massi04}. This extended jet-like radio structure, of angular extension of 10--50~milliarcseconds, seemed to be precessing. New and recent high resolution radio images obtained with the VLBA
during a full orbital cycle show an elongated structure changing its orientation at different orbital phases \cite{dhawan06}. This behaviour accommodates well in a model based in the interaction of a pulsar wind with
the dense equatorial wind from the Be star \cite{dubus06}. However, according to Ref.~\refcite{romero07}, this scenario presents some difficulties to explain the GeV luminosity of this source because the required spin-down power of the putative pulsar should be so high that the wind of the Be star could not collimate the radio emitting particles.

\subsection{LS~5039}
\begin{figure}[t]
\center
\centerline{\psfig{figure=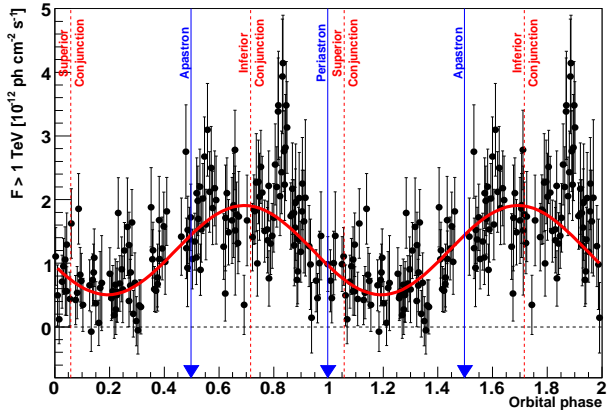,angle=0, width=8truecm}}
\caption{HESS integral $\gamma $-ray flux of LS~5039 as a function of orbital phase \cite{aharonian06}.\label{fig:ls}}
\end{figure}

LS 5039 is a HMXB system composed of a compact object of unknown mass, 
1.4--5$~M_{\odot}$
depending on the binary system inclination, that orbits an O6.5 main sequence star every 
3.9 d and is located
at 2.5 kpc \cite{casares05b}. Several 5 GHz VLBI observations of the source have been 
conducted in the
past with the VLBA+VLA, EVN and MERLIN, always showing a central core and bipolar radio jets 
\cite{paredes00,paredes02}. This led the authors to suggest that the system was a new microquasar in the Galaxy. LS 5039 was proposed
to be the counterpart of an unidentified EGRET source at HE $\gamma$-rays 
\cite{paredes00}. This has been confirmed after the detection in 
VHE gamma-rays (TeV energies) by the HESS Cherenkov telescope \cite{aharonian05b}, showing also a 3.9 d periodicity  (see Fig.~\ref{fig:ls}) \cite{aharonian06}. 

The differential photon energy
spectrum is variable with orbital phase. During the phases of the compact object inferior conjunction the spectrum is consistent
with a hard power-law where $\Gamma_{\rm VHE}$ = $1.85\pm0.06_{\rm stat}\pm0.1_{\rm syst}$ with exponential cutoff at $E_0$ = $8.7\pm2.0$ TeV. At the superior conjunction phases, the spectrum is consistent with a relatively
steep ($\Gamma_{\rm VHE}$ = $2.53\pm0.07_{\rm stat}\pm0.1_{\rm syst}$) pure power-law (0.2
to 10 TeV).
The HE/VHE emission is basically interpreted as the result of inverse Compton upscattering of
 stellar
UV photons by relativistic electrons. However, the nature of the acceleration mechanism that 
powers the
HE/VHE radiation is unclear, and two excluding scenarios have been proposed. In the first one
 electrons
are accelerated in the jet of a microquasar powered by accretion \cite{paredes06}. In the
 second one
they are accelerated in the shock between the relativistic wind of a non-accreting pulsar 
and the wind of the stellar companion \cite{dubus06}. In the pulsar scenario, a new model assuming two different sets of power-law spectral parameters of the interacting primary leptons has been developed to explain the HESS phenomenology \cite{sierpowska07}. Some important clues to understand the physics behind the TeV emission in LS 5039 have been recently presented in a scenario-independent work \cite{khangulyan08}. 

\begin{figure*}[t]
\center
\centerline{\psfig{figure=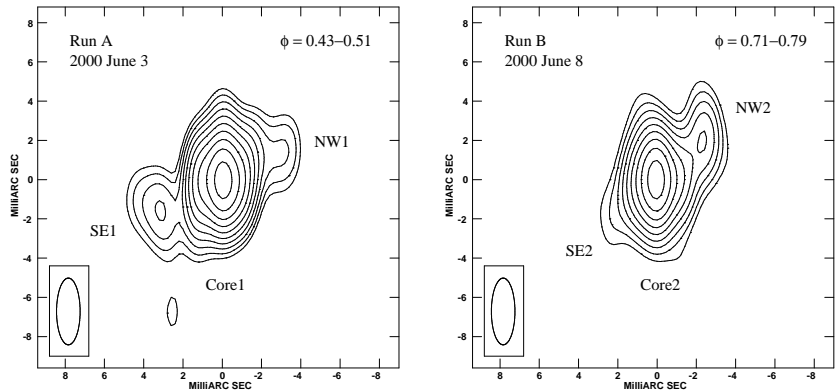, width=11truecm}}
\caption{VLBA maps of LS~5039 at different orbital phases \cite{ribo08}.\label{fig:lsvlbi}}
\end{figure*}

VLBI observations tracing the motion of 
the radio core and characterizing the changing morphology 
of LS 5039 during a complete orbital cycle would allow to unveil the nature of the particle 
acceleration mechanism and constrain some physical parameters, like the 
relativistic jet bulk velocity and inclination in the microquasar scenario or the binary 
system inclination in the non-accreting pulsar scenario. In a recent paper,\cite{ribo08} new VLBA radio observations of LS~5039 are presented. The imaging results (see Fig.~\ref{fig:lsvlbi}) 
show a changing morphology between two images obtained five days apart, whereas the 
astrometric results are not conclusive. This precludes to exclude clearly any of the 
two scenarios. While the observed changes in morphology cannot be 
explained easily with a simple and shock-less microquasar scenario, the young non-accreting 
pulsar scenario requires the inclination of the binary system to be very close to the upper 
limit imposed by the absence of X-ray eclipses.

\subsection{Cygnus~X-1}
Cygnus~X-1 is the first binary system where dynamic evidence for a BH
was found \cite{gies86}. According to the most recent estimates,
the BH mass is 20$\pm$5~$M_{\odot}$, while the O9.7\,Iab supergiant companion
has a mass of 40$\pm$10~$M_{\odot}$ \cite{ziolkowski05}. The orbit of the
system is circular, with a period of 5.6 days and an inclination of
33$\pm$5$^{\circ}$ \cite{gies86}. Located at 2.15$\pm$0.20~kpc
(3$\sigma$ error; see Ref.~\refcite{ziolkowski05} and references therein), Cygnus~X-1 is
the brightest persistent HMXB in the Galaxy, radiating a
maximum X-ray luminosity of a few times 10$^{37}$ erg~s$^{-1}$ in the 1--10~keV
range. 

The source displays the typical low/hard and high/soft states of accreting BH
binaries, spending most of the time, currently about 65\%, in the low/hard
X-ray state \cite{wilms06}. Steady compact jets are produced in BH
binaries in this state, when the inner radius of the disk is thought to be
truncated, while in the high/soft state the jet is quenched \cite{fender04}. This is also the case for Cygnus~X-1, which displays a $\sim$15~mJy and
flat spectrum relativistic compact (and one-sided) jet ($v>0.6c$) during the
low/hard state \cite{stirling01}, transient relativistic jets ($v$\simgt
$0.3c$) during state transitions \cite{fender06b}, while no radio emission
is detected during the high/soft state. 

Arc-minute extended radio emission around 
Cygnus X-1 was found \cite{marti96} using the VLA. Their disposition reminded 
of an elliptical ring-like shell with Cygnus X-1 offset from the center. 
Later, as reported in Ref.~\refcite{gallo05}, 
such structure was recognised as a
jet-blown ring around Cygnus~X-1. In analogy
with extragalactic jet sources, the ring could be the result of a strong shock
that develops at the location where the pressure exerted by the collimated jet,
detected at milliarcsec scales, is balanced by the ISM. If this pressure were solely due to a magnetized plasma in equipartition then we should expect a 
lobe synchrotron surface brightness much higher than the observed upper limit.  This suggest that most of the energy is stored in non-radiating particles,
presumably baryons.

The instrument COMPTEL, on board the {\it CGRO}, did detect Cygnus~X-1 in the
1--30~MeV range several times \cite{mcconnell00}. Unfortunately, EGRET
performed rather few observations of Cygnus~X-1 and did no detect the source.
Only a quite loose upper limit to the $\sim$100~MeV flux from Cygnus~X-1 is
available \cite{hartman99}.

\begin{figure}[t]
\center
\centerline{\psfig{figure=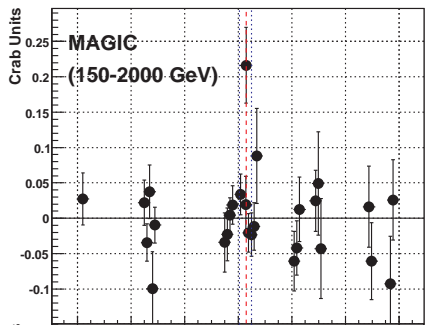, angle=0, width=5.6truecm}}
\caption{MAGIC measured fluxes from Cygnus~X-1 as a function of time \cite{albert07}. The minor thicks in the X-axis are separated five days.\label{fig:cygx1}}
\end{figure}

MAGIC observed Cygnus~X-1 in 2006 obtaining evidence of $\gamma$-ray signal with a significance of 4.9 $\sigma$ (4.1 $\sigma$ after trial correction)
\cite{albert07}. The signal was variable and extending in a short time interval ($\sim$ 80 minutes), as can be seen in Fig.~\ref{fig:cygx1}.
The measured 
excess is compatible with a point-like source at the position of 
Cygnus X-1 and excludes the nearby radio nebula powered by its relativistic jet.
 
 The differential energy spectrum is well fitted by a power law given
as $dN/(dA dt dE) = (2.3\pm0.6)\times10^{-12}(E/$1 TeV)$^{-3.2\pm0.6}$.
The measured excess was observed at phase 0.91, very near the inferior conjunction of the optical companion and
superior conjunction of the compact object \cite{gies03}. According to existing models dealing with photon-photon absorption and
cascading, there should not be detectable TeV emission at these orbital phases
if its origin is close to the compact object \cite{bednarek07} (behind the companion star in this configuration).

Interestingly, this detection occurred at the time when a flare was detected \cite{turler06} at
hard X-rays by {\it INTEGRAL} (at a level of about 1.5~Crab (20--40~keV) and
1.8~Crab (40--80~keV)) and {\it Swift}/BAT. The MAGIC detection occurred during a
particularly bright and hard X-ray flare that took place within a prolonged
low/hard state of Cygnus~X-1. The TeV peak appears to
precede a hard X-ray peak, while there is no particular change in soft X-rays.

\section{Scenarios for HE/VHE $\gamma$-ray production}\label{scenarios}
The observational results in the field of point-like TeV emission from galactic sources 
discussed above show that very efficient particle acceleration takes place in these compact 
sources. Photons have been observed up to energies as high as 30 TeV. 

Different works published during the last years have been based in the microquasar scenario \cite{paredes06,dermer06,bosch06}.  In these models, it has been proposed that the HE and VHE $\gamma$-ray emission of XBs with a jet (microquasars) is due to inverse Compton up-scattering of stellar UV photons from the companion star by the relativistic electrons of the jets.
 
Alternative models based on hadronic processes within the jets of a microquasar, or in the 
interaction between the relativistic jets and the interstellar medium (ISM), have also been 
proposed \cite{romero05}. Finally, a different scenario for the origin of HE and 
VHE $\gamma$-rays in these sources is the region of interaction between the relativistic wind 
of a young non-accreting neutron star and the wind of the stellar companion \cite{dubus06}. 

These models 
predict different multi-wavelength behaviors that can be probed observationally. In 
particular, very-high resolution interferometric radio observations (VLBI) can provide a 
final answer to this open question.

\section{Discussion and Conclusions}\label{discussion}

Up to now, 4 HMXBs but no LMXBs have been detected at TeV energies. This fact points the importance of having a bright companion (O or B star) as source of seed photons for the IC emission and target nuclei for hadronic interactions. The compact object in these binaries is in one case a BH and in another case a non-accreting NS, whereas there are two cases where the nature of the compact object is unknown.

Periodic TeV emission is present in one system, LS~5039, and likely in other two, \lsi\ and PSR~B1259$-$63. In Cygnus~X-1 the detected TeV emission comes from a flaring episode and the detection of a possible steady VHE flux is below the present IACT's sensitivity. The TeV flaring activity observed in Cygnus~X-1 might not be a unique and rare case. When looking at the integral $\gamma$-ray flux of LS~5039 (see Fig.~\ref{fig:ls}) it seems that at around phase 0.8 there is some flare activity superposed to the periodic-regular light-curve. In the case of \lsi\, in Fig.~\ref{fig:LSI} we can see the expected maximum at phase 0.6 and, in addition, a flaring activity at phase 0.8. This phenomenology seems to suggest that flare TeV emission might be common in XBs, although more flare detections are needed to establish this kind of TeV activity. 
                              
The increase of observational capability at VHE produced by VERITAS and CANGAROO III and the gain in sensitivity of HESS~II and MAGIC~II, will allow to detect new XBs and establish if the flaring activity is a common phenomena in these type of sources. All of this will bring more constraints to the physics of these systems. In addition, multi-wavelength (multi-particle) campaigns will also be very important to gain knowledge about these new TeV sources.

\section*{Acknowledgments}
The author acknowledges support of the Spanish Ministerio de Educaci\'on y Ciencia (MEC) under grant AYA2007-68034-C03-01 and FEDER funds. 


\balance

\end{document}